\documentclass[aps,prl,twocolumn,superscriptaddress,showpacs,floatfix,longbibliography]{revtex4-2}
\usepackage{amsmath,amssymb,amsfonts,float,graphics,epsfig,epstopdf,color,verbatim,tabularx,bm,multirow,appendix}
\usepackage[utf8]{inputenc}
\usepackage[T1]{fontenc}
\usepackage{xcolor}
\usepackage{dsfont}
\usepackage{textcomp}
\usepackage{yfonts}
\usepackage{footnote}
\usepackage{bm}
\usepackage{subfigure}
\usepackage{mathrsfs}
\usepackage{graphicx}
\usepackage{verbatim}
\usepackage{hyperref}
\usepackage{multirow}
\usepackage{braket}
\usepackage[normalem]{ulem}
\usepackage{tikz}
\usetikzlibrary{calc}
\usetikzlibrary{shapes.multipart}
\usepackage{young}
\usepackage{youngtab}
\usepackage{ytableau}

\renewcommand{\vec}[1]{{\mathbf #1}}

\newcommand{\comments}[1]{}

\newcommand{\hsbc}[1]{\hspace{4mm}#1\hspace{4mm}}

\newcommand{\Eq}[1]{Eq.~\eqref{#1}}

\newcommand{\Fig}[1]{Fig.~\ref{#1}}

\newcommand{\Tab}[1]{Tab.~\ref{#1}}

\newcommand{\stkout}[1]{\ifmmode\text{\sout{\ensuremath{#1}}}\else\sout{#1}\fi}

\newcommand\startsupplement{%
       \newpage\clearpage
       \setcounter{secnumdepth}{2}
       \setcounter{table}{0}
       \renewcommand{\thetable}{S\arabic{table}}
       \setcounter{figure}{0}
       \renewcommand{\thefigure}{S\arabic{figure}}
       \setcounter{equation}{0}
       \renewcommand{\theequation}{S\arabic{equation}}
       \setcounter{section}{0}
       \renewcommand{\thesection}{Section \Roman{section}}
       \renewcommand{\thesubsection}{\Roman{section}. \Alph{subsection}}
    }

\makeatletter
\def\l@subsubsection#1#2{}
\makeatother

\graphicspath{{./Figs/}}

\begin{document}

\title{Emergent Conformal Symmetry at the Multicritical Point of (2+1)D SO(5) Model with Wess-Zumino-Witten Term on Sphere}

\author{Bin-Bin Chen}
\affiliation{Department of Physics and HKU-UCAS Joint Institute of Theoretical and Computational Physics, The University of Hong Kong, Pokfulam Road, Hong Kong SAR, China}

\author{Xu Zhang}
\affiliation{Department of Physics and HKU-UCAS Joint Institute of Theoretical and Computational Physics, The University of Hong Kong, Pokfulam Road, Hong Kong SAR, China}

\author{Zi Yang Meng}
\email{zymeng@hku.hk}
\affiliation{Department of Physics and HKU-UCAS Joint Institute of Theoretical and Computational Physics, The University of Hong Kong, Pokfulam Road, Hong Kong SAR, China}

\begin{abstract}

Novel critical phenomena beyond the Landau-Ginzburg-Wilson paradigm have been long sought after.
Among many candidate scenarios, 
the deconfined quantum critical point (DQCP) constitutes the most fascinating one, and its lattice model realization has been debated over the past two decades. 
Following the pioneering works with the fuzzy sphere methods~\cite{zhuUncovering2023,huOperator2023,hanConformal2023,zhouThe2023}, 
we apply the spherical Landau level regularization to study the effective 
(2+1)D SO(5) non-linear sigma model with a topological term and the potential DQCP therein. 
Utilizing the state-of-the-art density matrix renormalization group method with 
explicit {$\text{SU(2)}_\text{spin}\times\text{U(1)}_\text{charge}\times\text{U(1)}_\text{angular-momentum}$} symmetry as well as exact diagonalization simulations, we provide a comprehensive
phase diagram for the model with a SO(5) continuous transition line --- extension of the previous identified SO(5) multicritical point~\cite{chenPhases2023} --- while tuning interaction length. 
The state-operator correspondence with the conformal tower structure is used to identify the emergent conformal symmetry with the best scaling dimension of relevant primary fields and they match well with the critical exponents obtained from the crossing point analysis of the correlation ratio. Our results thus  further support the rich structure of the phase diagram of the SO(5) model. 
\end{abstract}

\date{\today}
\maketitle

\noindent{\textcolor{blue}{\it Introduction.}---}
Novel critical phenomena beyond Landau-Ginzburg-Wilson paradigm can fundamentally enrich our understanding to the phase transitions of highly entangled quantum matter. Amongst many, the deconfined quantum criticality, known as the direct continuous transition between two spontaneous symmetry-broken states, has been long-sought-after~\cite{senthilQuantum2004,wangDeconfined2017,qinDuality2017,maDynamics2018,senthilDeconfined2023}. However, the initial proposed lattice realizations~\cite{sandvikEvidence2007,louvbsneel2009}, i.e. from N\'eel to valence bond solid (VBS) transition in $J$-$Q$ spin model, with drifting critical exponents are incompatible with conformal bootstrap bounds~\cite{haradaPossibility2013,nahumDeconfined2015,shaoQuantum2016,nakayamaNecessary2016,polandConformal2019,li2022bootstrapping}, has been shown to exhibit weakly first-order transition behavior, over the years of heroic efforts~\cite{kuklovDeconfined2008,haradaPossibility2013,chenDeconfined2013,nahumDeconfined2015,zhaoMulticritical2020,takahashi2024so5}.
%\bc{Furthermore, the drifting critical exponents are incompatible with conformal bootstrap bounds~\cite{haradaPossibility2013,nahumDeconfined2015,shaoQuantum2016,nakayamaNecessary2016,polandConformal2019,li2022bootstrapping}.}
 It can turn into a continuous transition---a real DQCP---at least when $N>8$ in the SU($N$) version of $J_1$-$J_2$ spin model~\cite{kaulLattice2012,blockFate2013,songExtracting2023,song2023deconfined}. At the large $N$ limit, the transition is believed to be described by the Abelian Higgs theory with unitary conformal fixed point~\cite{irkhinExpansion1996,sachdevQuantum2008,ihrigAbelian2019}.

The obstacles of finding a DQCP in the realistic SU(2) setting originate from the symmetry emergence requirement. To be specific, in the SU(2) version of $J$-$Q$ model on square lattice, a U(1) symmetry is firstly required to emerge from the $\mathbb{Z}_4$ symmetry in the VBS phase around the transition point~\cite{sandvikEvidence2007}. 
However, such U(1) symmetry emergence is subtle due to the dangerously irrelevant pertubation (the monopole event)~\cite{shaoQuantum2016,polandConformal2019}:  the U(1) is emergent only in the infrared limit, while in contrast the $\mathbb{Z}_4$ symmetry breaking term is relevant, which obscures the numerical 
analysis when an additional $\mathbb{Z}_4$ symmetry length scale is approached.
Furthermore, the emerged U(1) symmetry in the VBS phase should again combine with the SU(2) symmetry in the Néel phase to give rise to a final SO(5) symmetry~\cite{nahumEmergent2015}. Such symmetry emergence requirement might set up a high bar for realizing SO(5) in lattice models, where a slow renormalization group (RG) flow may cause considerable finite-size effect, and eventually lead to the observation of the weakly first order transition after years of accumulating works.

Given the situation, an explicit SO(5) model with Wess-Zumino-Witten (WZW) topological term by projecting fermion density-density interaction on the Landau level is proposed to circumvent such symmetry emergence requirement in the N\'eel-VBS lattice models~\cite{leeWess2015}. Such a model has been numerically visited on both torus~\cite{impoliteHalf2018,wangPhases2021} and recently spherical geometries~\cite{zhouThe2023,chenPhases2023}. 
{As pointed out in the seminal works of Refs.~\cite{zhuUncovering2023,huOperator2023,hanConformal2023,zhouThe2023}}, 
the advantage of the spherical geometry
is to directly expose the underlying conformal field theory (CFT) algebra and operator content, since
for a Hamiltonian living on $S^{d-1} \times R$ space-time geometry, 
the scaling dimensions of CFT operators has one-to-one correspondence with 
the eigenenergies of CFT states, dubbed as the state-operator correspondence~\cite{DiFrancesco_CFT,Rychkov_2017,cardyOperator1986}.
Recently, by using the idea of the spherical regularization~\cite{haldaneFractional1983,Arciniaga2016}, 
the conformal data including the scaling dimensions and operator product expansion (OPE) 
coefficients have been characterised in the (2+1)D Ising~\cite{zhuUncovering2023,huOperator2023}, 
O(3) critical points~\cite{hanConformal2023}, {and SO(5) pseudo-criticality~\cite{zhouThe2023}} 
(we also note an conformal perturbation theory approach in Ising~\cite{Lao2023ConformalPerturbation} and O(2) cases~\cite{herviou2024}).
{The conformal defect can also be studied with the fuzzy sphere regularization~\cite{hu2023solving,zhou2024gfunction}.}

In our previous work of SO(5) model on sphere~\cite{chenPhases2023}, we mapped out the phase diagram of the model, not only along the SO(5) symmetric line but also break the symmetry down to SO(3)$\times$SO(2), and probe all the possible symmetry-breaking phases therein. 
We find a possibly gapless SO(5) symmetric phase separating the N\'eel and VBS phases, and the  
transitions from the symmetric phase towards N\'eel and VBS symmetry breaking phases acquire non-Wilson-Fisher critical exponents. 
More importantly, these two phase boundaries meet at a multicritical point below which the SO(5) symmetry is spontaneously broken and the Néel-VBS transition becomes first-order. Our results are consistent with recent understanding that the aforementioned N\'eel-VBS weakly first-order transition~\cite{kuklovDeconfined2008,haradaPossibility2013,chenDeconfined2013,nahumDeconfined2015,zhaoMulticritical2020,songExtracting2023,song2023deconfined,takahashi2024so5} is located close but below our multicritical point, and the recent conformal bootstrap analysis of deconfined quantum tricriticality~\cite{chesterBootstrapping2024}. However, the precise critical exponents are hard to derive there because of the still relatively large finite size effect~\cite{zhouThe2023,chenPhases2023}. One would need to further fine-tune the Hamiltonian (preserving SO(5) symmetry) to fully review the emergent conformal symmetry structure at the multicritical point within reachable sizes of such spherical setting.

\begin{figure}[t!]
\includegraphics[width=\columnwidth]{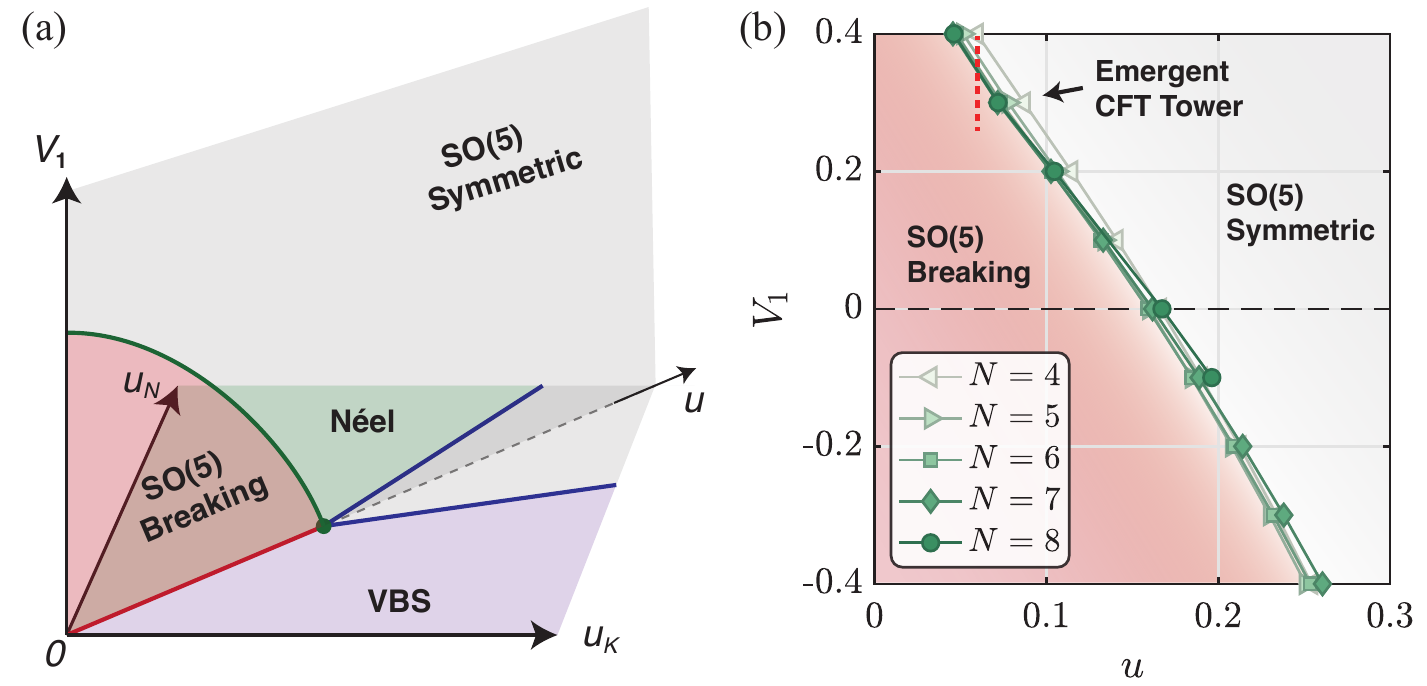}
\caption{\textbf{Global phase diagram of the SO(5) model.}  (a) The $u_K$-$u_N$ plane is obtained from our previous work~\cite{chenPhases2023}, where tuning along the $u$ axis respects the SO(5) symmetry. We found a multicritical point (the green dot) separating the SO(5) symetry-breaking line (the red segment) and a SO(5) symmetric phase which further separates the N\'eel (green) and VBS (blue) phases. In this work, we extend the interaction to longer range by adding the $V_1$ term in the Hamiltonian (preserve the SO(5) symmetry), which allows us to scan the $u$-$V_1$ plane to look for the transition point between SO(5) breaking and SO(5) symmetric phases with the emergent CFT structure.
(b) Obtained phase boundary in the $u$-$V_1$ plane by identifying the calibrated scaling dimension 
$\Delta_{\mathcal{T}^{\mu\nu}}=3$. One sees the boundary is converged with $N=6$ near the black arrow where the emergent CFT structure is discovered close to $V_1=0.3$.
The red dashed line indicates the DMRG simulations in \Fig{fig:fig4} where the critical point and scaling dimension from finite size crossing point analysis are consistent with the CFT spectrum results.  
}
\label{fig:fig1}
\end{figure}

In this work, we accomplish this goal by extending the interaction length of the Hamiltonian, 
which allows us to search for a group of parameters with smaller finite-size effect in the enlarged parameter space, 
where irrelevant operators are sufficiently suppressed. 
{Such prescription has been successfully applied to the studies in the 3D Ising and O(3) transitions in 
fuzzy sphere method~\cite{zhuUncovering2023,huOperator2023,hanConformal2023}.} 
{With that parameters, several low-lying primary operators show perfectly 
conformal tower structure indicating a CFT here. 
The phase transition point and scaling dimension of order parameter 
from finite-size crossing point analysis of the SO(5) correlation ratio~\cite{qinDuality2017,shaoQuantum2016,maAnomalous2018,maRole2019,luckCorrections1985,chenPhases2023},
match well with the conformal tower results. 
This self-consistence check indicates the robustness of our conclusion.}
Our results {provide a strong evidence for the conformal symmetry of 
the multicritical SO(5) CFT fixed point}, 
which separates the SO(5) symmetry-breaking first-order transition line and the SO(5) symmetric phase 
in the global phase diagram of the model (Fig.~\ref{fig:fig1}), 
and it is from this multicritical point the non-Wilson-Fisher phase boundaries 
between the symmetric to N\'eel and VBS phases are originated.

\begin{figure}[t!]
	\includegraphics[width=\columnwidth]{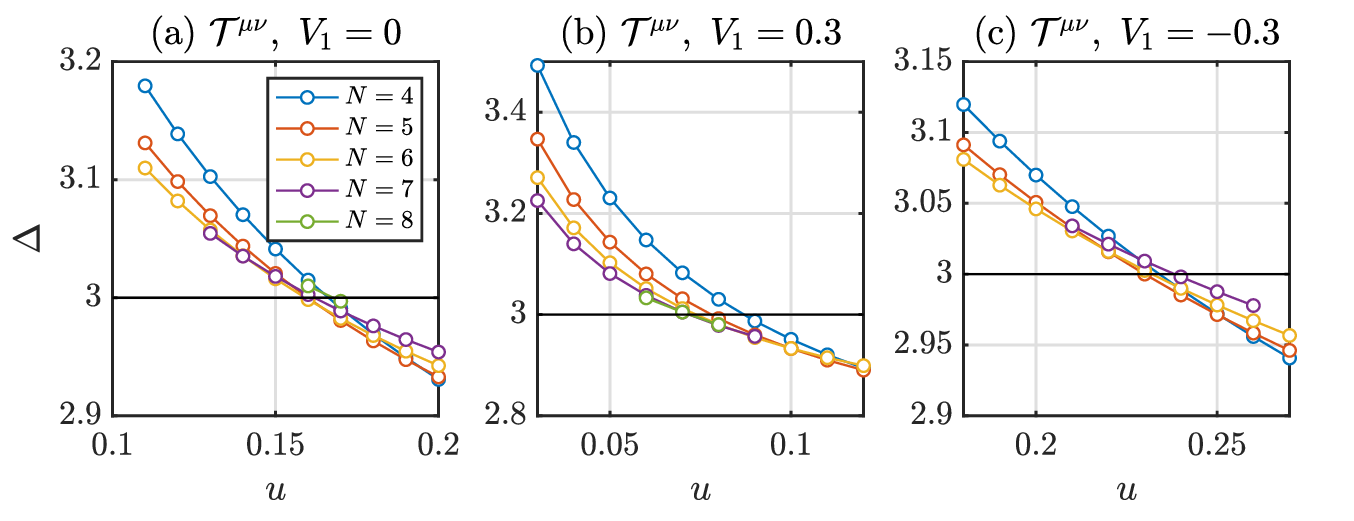}
	\caption{\textbf{Scaling dimension of stress tensor $\Delta_{\mathcal{T}^{\mu\nu}}$}. 
	Calibrated by $\Delta_{J^\mu}=2$, 
	$\Delta_{\mathcal{T}^{\mu\nu}}$'s are shown versus $u/U_0$ for different system sizes 
	$N = 4,5,6,7,8$, for different interaction strength 
	(a) $V_1 = 0$, (b) $V_1 =0.3$, and (c) $V_1=-0.3$. The $\Delta_{\mathcal{T}^{\mu\nu}}=3$ criterion determines the transition point for each $V_1$.
	}
	\label{fig:fig2}
\end{figure}

\noindent{\textcolor{blue}{\it Model and Methods.}---}
We consider the (2+1)D %SO(5) non-linear sigma model 
Hamiltonian
$H_\Gamma = \int d\mathbf{r}_1 d\mathbf{r}_2 U(\mathbf{r}_1,\mathbf{r}_2) 
[n^0(\mathbf{r}_1)n^0(\mathbf{r}_2)-\sum_{i=1}^5 u_i~ n^i(\mathbf{r}_1)n^i(\mathbf{r}_2)]$, 
where $n^i(\mathbf{r})={\bf\Psi(r)}^\dag\Gamma^i{\bf\Psi(r)}-2\delta_{i0}$ is a local density operator with 
${\bf\Psi(r)}\equiv\psi_{\tau\sigma}(\mathbf{r})$ the 
4-component Dirac fermion annihilation operator with 
mixing valley $\tau$ and spin $\sigma$ indices. 
Here $\Gamma^0=\mathbb{I}\otimes\mathbb{I}$ is the identity matrix 
and $\{\Gamma^{1,\cdots,5}\}=
\{\tau_x\otimes\mathbb{I}, \tau_y\otimes\mathbb{I}, \tau_z\otimes{\sigma_x}, 
\tau_z\otimes{\sigma_y}, \tau_z\otimes{\sigma_z}\}$ are the 5 mutually 
anticommuting matrices of the SO(5) group.
The short-range density-density interaction is
$U(\mathbf{r}_1,\mathbf{r}_2) = \frac{g_0}{R^2}\delta(|\mathbf{r}_1-\mathbf{r}_2|)+
\frac{g_1}{R^4}\nabla^2\delta(|\mathbf{r}_1-\mathbf{r}_2|)$ where $g_1$ is used to tune the interaction length. In previous works~\cite{zhouThe2023,chenPhases2023}, only a purely local interaction is considered with $g_1=0$.

Subsequently, we project the SO(5) Dirac fermion Hamiltonian onto the zero energy Landau level on the sphere, which is the same as the massive fermion lowest Landau levels (LLL) of a sphere with $4\pi s$ magnetic monopole at its origin~\cite{haldaneFractional1983,JELLAL2008,Arciniaga2016,GREITER2018}, where the $(2s+1)$-fold degenerate 
LLL wavefunction takes the form of 
$\Phi_m(\Omega)\propto e^{im\phi}\cos^{s+m}(\tfrac{\theta}{2})\sin^{s-m}(\tfrac{\theta}{2})$ 
with $m\in\{-s, -s+1,\cdots,s-1,s\}$ and $2s \in\mathbb{Z}$. By enlarging $s$ one can effectively enlarge the surface (system size) of the sphere while keeping the local magnetic field on the surface unchanged. After projection $\psi(\Omega)\rightarrow\sum_m \Phi_m(\Omega)c_m$, one can derive the projected Hamiltonian 
\begin{align}\label{Eq:Model}
\begin{aligned}
&\hat H_\Gamma^{(LLL)} = \hat H_0 - \sum_{i=1}^{5}u_i~ \hat H_i, \text{with} \\
&\hat H_{i\in\{0,1,...,5\}}=\sum_{m_1,m_2,m}V_{m_1,m_2,m_2-m,m_1+m}\times\\
& \left(c^\dag_{m_1}\Gamma^i c^{\,}_{m_1+m}-2\delta_{i0}\delta_{m0}\right)
 \left(c^\dag_{m_2}\Gamma^i c^{\,}_{m_2-m}-2\delta_{i0}\delta_{m0}\right)
 \end{aligned}
\end{align}
where $V_{m_1,m_2,m_3,m_4}$ is connected to the 
Haldane pseudo-potential $V_l$ by 
$V_{m_1,m_2,m_3,m_4} = \sum_l V_l (4s-2l+1)
\begin{pmatrix}
s & s & 2s-l \\
m_1 & m_2 & -m_1-m_2
\end{pmatrix}
\begin{pmatrix}
s & s & 2s-l \\
m_4 & m_3 & -m_3-m_4
\end{pmatrix}$. 
For the considered short-range interaction here, only $V_0$ and $V_1$ are involved in the
above $l$-summation [c.f. Supplemental Material (SM) for the relation between $(g_0, g_1)$ and $(V_0, V_1)$~\cite{suppl}].
This model is known to be described by a SO(5) non-linear sigma model with a WZW term~\cite{leeWess2015,impoliteHalf2018,wangPhases2021}.
We define $u_K=u_1=u_2$ for the VBS control parameter, and $u_N=u_3=u_4=u_5$ for the N\'eel
control parameter. 
Throughout this work, we set $V_0=1$ as the energy unit and focus on the 
SO(5) case, i.e. $u=u_K=u_N$ while tuning 
$u$ (distance from SU(4) fixed point) and $V_1$ (finite-length interaction strength) to reduce 
the finite size effect of the SO(5) phase boundary found in our previous work~\cite{chenPhases2023}. Details of the spherical regulation is given in SM~\cite{suppl}. {We note this approach is used in Ref.~\cite{zhouThe2023} to propose the pseudo-criticality along the $u_K=u_N$ SO(5) line.}

In this work, we employ density matrix renormalization group (DMRG) method with 
$\text{SU(2)}_{\text{spin}}\times \text{U(1)}_{\text{charge}}\times \text{U(1)}_{\text{angular-momentum}}$ symmetry, and exact diagonalization (ED), to accurately determine the phases and their phase boundaries. 
In our DMRG simulations, the SU(2) symmetry is implemented in the framework of the
tensor library QSpace~\cite{Weichselbaum2012,Weichselbaum2020,Bruognolo2021},
with up to $4096$ SU(2) invariant multiplets (equivalent to $\sim12000$ 
U(1) states) kept to render the truncation errors within $5\times10^{-5}$.
We denote the system size {by the Landau level degeneracy} $N=2s+1$ and {obtain converging} 
results with $N=3,4,5,...,{12}$
to control finite size scaling behavior.
In the energy level computation, we exploit ED for $N=4,5,\cdots,7$ and DMRG for $N=8$.

\noindent{\textcolor{blue}{\it Phase Diagram and the SO(5) phase transition.}---}
We first give a summary of the phase diagram. 
As shown in \Fig{fig:fig1}~(a), for the case of $V_1=0$, the $u_K-u_N$ plane of the phase diagram 
is investigated in our previous work~\cite{chenPhases2023}, where generically the SO(5) symmetry 
is split into SO(3)$\times$SO(2) symmetry and the N\'eel (green) and VBS (purple) phases 
are either separated by a direct first-order transition (solid red line) 
or through an intermediate symmetric phase (grey area). 
The SO(3) and SO(2) transition lines (solid blue line) merge into a fine-tuned 
multi-critical point (green dots) in the plane.
Here, we will show that, in the SO(5) cases of $u_K=u_N=u$, by adding a $V_1$ axis, 
the multi-critical point further extends as a transition line separating the SO(5)-breaking 
(red area) and the SO(5)-symmetric phases.

To identify such SO(5) transition line, we resort to  
the state-operator correspondence, i.e.,  
for a CFT operator $O_k$, the corresponding energy gaps $E_k-E_0$ are 
proportional to the scaling dimensions $\Delta_k$, 
i.e., $E_k-E_0 = \frac{v}{R} \Delta_k$, where $R$ is the radius of the sphere and 
$v$ is a model-dependent velocity~\cite{cardy_conformal_1984,cardy_epidemic_1985}. 
To search for the potential SO(5) CFT points in the $V_1$-$u$ phase diagram, 
we need to firstly match the energy spectra of the considered SO(5) model with 
the operator spectrum, to be specific, determining the velocity $v$ and to rescale 
the energy spectra with the factor $v/R$. Note that, the symmetry 
current $J^\mu$ and the energy-momentum tensor $\mathcal{T}^{\mu\nu}$ don't renormalize 
in perturbation theory, i.e. they have zero anomalous dimesions, meaning that 
their scaling dimensions $\Delta_{J^\mu}=d-1=2$ and 
$\Delta_{\mathcal{T}^{\mu\nu}}=d=3$~\cite{Rychkov_2017}, 
which can serve as natural calibrators for the operator spectrum. 

{With the prescription established in 3D Ising transition~\cite{zhuUncovering2023,huOperator2023}, 
O(3) Wilson-Fisher transition~\cite{hanConformal2023}, and SO(5) pseudo-criticality~\cite{zhouThe2023}}, 
we rescale the energy spectrum by exactly setting $\Delta_{J^\mu}=2$ and 
searching in the $V_1$-$u$ parameter space for $\Delta_{\mathcal{T}^{\mu\nu}}=3$. 
This will determine a critical line which should coincide with the SO(5) phase boundary 
obtained from the correlation ratio data, as we will discuss in \Fig{fig:fig4}. 
As shown in \Fig{fig:fig2}, at the fixed $V_1=0$, $0.3$, $-0.3$ cuts, we plot 
the scaling dimension of $\mathcal{T}^{\mu\nu}$ versus $u$ for various system sizes 
$N = 4,5,...,{8}$. 
By collecting those $\Delta_{\mathcal{T}^{\mu\nu}}=3$ points and we plot them 
as the phase boundary in \Fig{fig:fig1}~(b).

\begin{figure*}[t!]
\includegraphics[width=1.6\columnwidth]{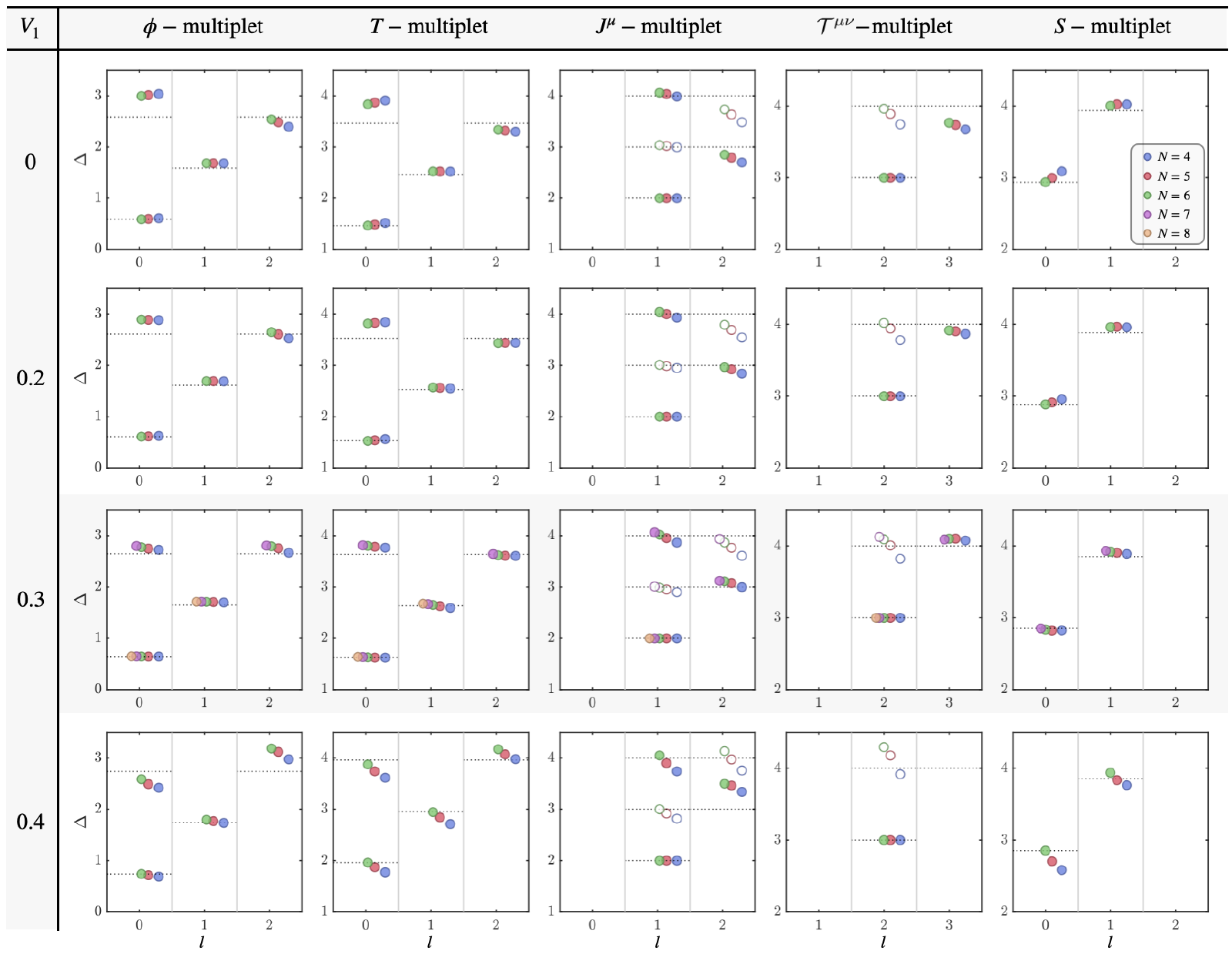}
\caption{\textbf{Conformal multiplets of several low-lying primary operators.} 
Scaling dimension $\Delta$ versus Lorentz spin $l$ for 
the lowest vector $\phi$ (1st column), 
the lowest rank-2 symmetric traceless tensor $T$ (2nd column),
the conserved current $J^\mu$ (3rd column), 
the stress tensor $\mathcal{T}^{\mu\nu}$ (4th column) and the scalar (singlet operator) 
$S$ (5th column), 
when tuning interacting $V_1 = 0$ (1st row), 
$V_1 = 0.2$ (2nd row), 
$V_1 = 0.3$ (3rd row), 
$V_1 = 0.4$ (4th row).
{Following the pioneering work of Ref.~~\cite{zhouThe2023},} the scaling dimensions $\Delta$'s are calibrated by the scaling dimension of 
the conserved current $\Delta_{J^\mu}=2$. We find the emergent CFT structure manifest close to $V_1=0.3$ (indicated by the grey background) and it represent the CFT of the SO(5) multicritical point. 
Here the solid/empty-filled circles depict the first/second type descendants 
[c.f. \Eq{Eq:CFTDescendantT1}, \Eq{Eq:CFTDescendantT2}].
{Note that, the first row ($V_1=0$) is consistent with the CFT spectrum observed in Ref.~\cite{zhouThe2023}.}
}
\label{fig:fig3}
\end{figure*}

\noindent{\textcolor{blue}{\it Energy spectrum evidence for conformal symmetry.}---}
The finger-print evidence for the CFT nature is the integer-spaced tower structure 
of the primary operators and their descendant's~\cite{zhuUncovering2023}. 
That is, for scalar primary operator 
$\hat O$ with scaling dimension $\Delta$ and Lorentz spin $l=0$, its descendants express
as $\partial_{\nu_1}\cdots\partial_{\nu_m}\Box^n \hat O$ with $m,n\geq0$ whose scaling 
dimension will be $\Delta+2n+m$, and the Lorentz spin will be $m$~\cite{DiFrancesco_CFT,Rychkov_2017}. 
Note that, for spin-$l$ primary operator $O_{\mu_1\cdots\mu_l}$, e.g. the spin-$1$ 
current operator $J^\mu$ and the spin-$2$ stress tensor operator $\mathcal{T}^{\mu\nu}$, 
there are two types of descendants. The first type is expressed as 
\begin{equation}\label{Eq:CFTDescendantT1}
\partial_{\nu_1}\cdots\partial_{\nu_m}\partial_{\mu_1}\cdots\partial_{\mu_i}\Box^n \hat O_{\mu_1\cdots\mu_l}
\end{equation}
whose scaling dimension will be $\Delta+2n+m+i$, and the Lorentz spin will be $l+m-i$. 
%\bc{According to Refs.~\cite{zhouThe2023,hanConformal2023},} 
The second type can be expressed as 
\begin{equation}\label{Eq:CFTDescendantT2}
\epsilon_{\mu_l\rho\tau}\partial_\rho\partial_{\nu_1}\cdots\partial_{\nu_m}\partial_{\mu_1}\cdots\partial_{\mu_i}\Box^n \hat O_{\mu_1\cdots\mu_l}
\end{equation} 
whose scaling dimension will be $\Delta+2n+m+i+1$, and the Lorentz spin will be $l+m-i$. 
The $\epsilon$ tensor of SO(3) will alter spacetime parity symmetry of $\hat O_{\mu_1\cdots\mu_l}$. 
In addition, for conserved operators like $J^\mu$ and $\mathcal{T}^{\mu\nu}$, due to the 
conservation law $\partial_\mu J^\mu = 0$ and $\partial_\mu \mathcal{T}^{\mu\nu}=0$, their
descendants should have $i=0$.

Due to finite size effect, 
the operator spectrum might lose such integer-space structure, and our second 
step is to suppress the irrelevant operators to expose the CFT tower in a small size simulation by tuning the parameter $V_1$ along the phase boundary, {as has been employed in 3D Ising and O(3) transitions~\cite{zhuUncovering2023,huOperator2023,hanConformal2023}}. Here, we 
consider the SO(5) order parameter $\phi$ (the lowest vector representation), 
the lowest rank-2 symmetric traceless tensor $T$, 
the symmetry current  $J^\mu$, the stress tensor $\mathcal{T}^{\mu\nu}$, 
the SO(5) singlet operator $S$ (the lowest scalar representation), 
as well as their descendant's. The degeneracy of these operators with their associated quantum numbers are explicitly given in the tensor representation of the SO(5) group in SM~\cite{suppl}.

As shown in \Fig{fig:fig3}, we align the towers of 
different operators in separated columns, i.e., $\phi$ the first, $T$ the second, 
$J^\mu$ the third, $\mathcal{T}^{\mu\nu}$ the fourth, and $S$ the last column. 
The different rows indicate different interactions strength $V_1=0, 0.2, 0.3, 0.4$. 
{We note, the first row ($V_1=0$) corresponds to the CFT spectrum observed in Ref.~\cite{zhouThe2023}.}
In each plot, the operator spectrum with different Lorentz spin $l$ is split by 
the vertical solid lines, and the horizontal dotted lines indicate the perfect 
integer-spaced tower structure. The numerical results for the first/second type descendants are depicted as 
solid/empty-filled circle in the plots.
It can be clearly seen that, as $V_1$ increases, 
the CFT towers get increasingly well-behaved, i.e., each operator approaches to the 
corresponding integer-spaced horizontal line. At $V_1=0.3$, we find, 
the operator spectra recover the CFT towers up to second descendant operators. {Further increase $V_1$ above 0.3, the structure disappear again. Therefore, we find close to $V_1=0.3$ and along the $u$ scan, the finite size effect, i.e. the irrelevant operators, has been successfully removed.}

\noindent{\textcolor{blue}{\it Scaling dimensions of relevant operators.}---}
Other than the above primary operators, i.e., $\phi, T, S$, we also find another 
relevant operator, the 6$\pi$-monopole operator $M_{6\pi}$ whose scaling dimension 
is smaller than $3$. As shown in \Tab{tab:tab1}, we list all the relevant operators 
at $V_1=0.3$. 
Remarkably, we find that, the singlet operator $S$ and the rank-2 tensor $T$ possesse scaling 
dimensions $\Delta_S\simeq2.884 <3$ and $\Delta_T\simeq1.636 <3$ which means that the 
operator away from the SO(5) axis is relevant. 
That is, the observed SO(5) CFT is a multicritical point 
(in the SO(3)$\times$SO(2) $u_K$-$u_N$ plane) with two relevant singlet operators, 
the SO(5) singlets $S$ and $T$ 
(this singlet comes from decomposing $T$ of SO(5) to SO(3)$\times$SO(2)~\cite{chesterBootstrapping2024}), 
in consistence with the recent conformal bootstrap analysis of the deconfined quantum 
tricritical scenario~\cite{chesterBootstrapping2024}. 
{We also note that, Ref.~\cite{zhouThe2023} obtained the similar dimensions 
$\Delta_S = 2.831$, $\Delta_T = 1.458$ while having different interpretation of 
pseudocriticality due to the vast region of CFT behaviour in the case of $V_1=0$, 
ruling out the possibility of a quantum transition therein. 
Here, as shown in the \Fig{fig:fig1}(b) that, the phase boundaries 
still vary slowly with system size $N$, which can be improved by increasing $V_1$ 
with well-converged phase boundaries around $V_1=0.3$, 
signaling the stabilization of the transition point and 
the validity of the fine-tuning multicriticality scenario.}

\begin{table}[tb]
\caption{Scaling dimension of relevant primary operator at $V_1=0.3$ for various system sizes $N$. 
}
\begin{tabular}{c|ccccc}
    \toprule
         & &  & $N$ &  &  \\
    \hline
    Operators &\hsbc{4} & \hsbc{5} & \hsbc{6} & \hsbc{7} & \hsbc{8} \\
    \hline
    $\phi$ & 0.642  & 0.642 & 0.644  & 0.646 & 0.647  \\ [2pt]      
    $T$  & 1.622 & 1.622 & 1.627  & 1.633 & 1.636   \\ [2pt]   
    $J^\mu$  & 2.000  & 2.000 & 2.000  & 2.000  & 2.000  \\ [2pt]   
    $S$  & 2.853  & 2.823 & 2.873 & 2.884 & ---   \\ [2pt]   
    $M_{6\pi}$  & 2.825  & 2.861 & 2.836  & 2.852 & ---   \\ [2pt]
    $\mathcal{T}^{\mu\nu}$  & 3.000  & 3.000 & 3.000  & 3.000  & 3.000  \\      
    \hline\hline
\end{tabular}
\label{tab:tab1}
\end{table}

\begin{figure}[t!]
\includegraphics[width=\columnwidth]{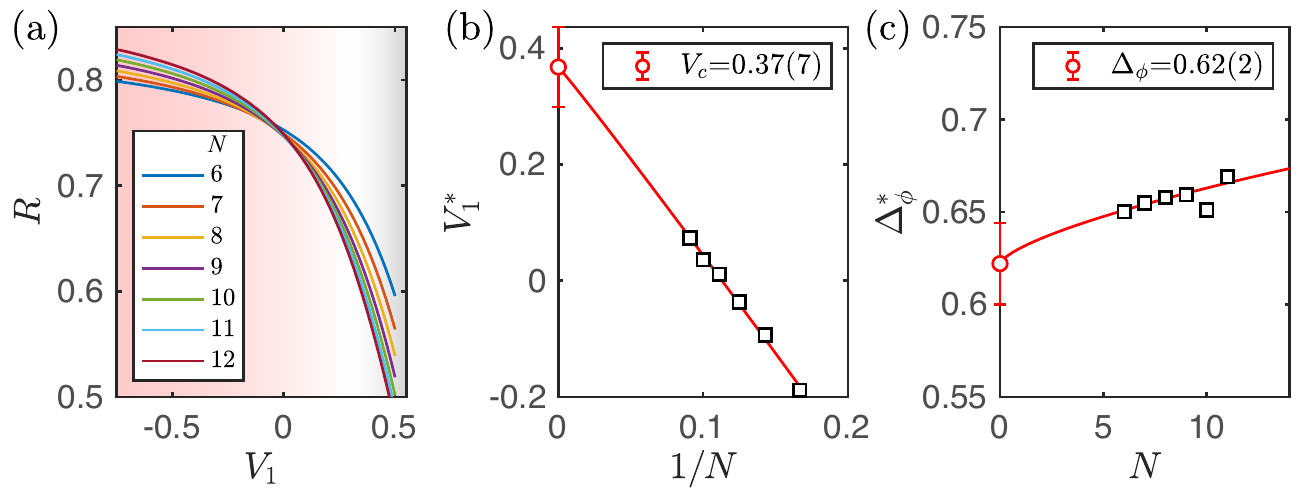}
\caption{\textbf{DMRG data for SO(5) transition.} 
(a) Correlation ratio $R$ of SO(5) order parameter are shown versus $V_1$ at fixed $u=0.06$.  
(b) The critical point is extrapolated from the finite-size crossing point $V_c^\ast$ and 
yields $V_c=0.37(7)$, consistent with the phase boundary in \Fig{fig:fig1} (b).
(c) The scaling dimension extrapolated to be $\Delta_\phi = 0.62(2)$ is also consistent with CFT spectrum in Tab.~\ref{tab:tab1}. }
\label{fig:fig4}
\end{figure}

As an important sanity check, we also determine the scaling dimension of 
the order parameter by calculating the correlation ratio of the SO(5) order parameter in DMRG~\cite{chenPhases2023}.
For the SO(5) ordered phase, we define 
$O_{i,l,m} = {\int{d\mathbf{r} ~Y_{lm}^\ast\left( \mathbf{r} \right)\mathbf{\Psi}^{\dagger}\left( \mathbf{r} \right)\Gamma^{i}\mathbf{\Psi}\left( \mathbf{r} \right)}}$ 
with $Y_{l,m}(\mathbf{r})$ being the spherical harmonic function, 
and compute the squared order parameter 
$m^2_l = \frac{1}{N^2}\sum_{i=1}^5\langle O_{i=1,l}^2\rangle$ 
and the corresponding correlation ratio 
$R = 1-m^2_{l=1}/m^2_{l=0}$.

In \Fig{fig:fig4}~(a), we fix $u=0.06$ and vary $V_1$. The VBS correlation ratio $R$ 
exhibits a nice crossing behaviour, which 
suggests the SO(5)-breaking phase at smaller $V_1$ and the symmetric phase at larger $V_1$. As in our previous work~\cite{chenPhases2023}, it can be seen that, the crossing points exhibit slightly drifting behaviour, which can 
be described by the scaling form 
$V_1^\ast(N,N+1) = V_{c} + N^{-\frac{1}{2\nu}-\frac{\omega}{2}}$ 
(the asterisk indicates the finite-size crossing points and 
the critical point in the thermodynamic limit is indicated by the subscript `$c$').
In \Fig{fig:fig4}~(b), we find the crossing points are best fitted by a linear form 
$V_1^\ast = a + b/N$, and obtain the 
critical point from the intercept $V_c=a=0.37(7)$, consistent with the phase diagram in \Fig{fig:fig1} (b). We then define 
$\Delta^\ast_\phi(N) = -N\log\frac{m^2(V_c,N+1)}{m^2(V_c,N)}$, 
which follows the scaling form $\Delta^\ast_\phi(N) = \Delta_\phi + a N^{\frac{1}{2\nu}}$. 
Here the exponent $\nu=\frac{1}{3-\Delta_T}\simeq0.733$ is determined from 
the $\Delta_T$ in Tab.~\ref{tab:tab1}. And as shown in \Fig{fig:fig4}~(c), 
the order parameter scaling dimension $\Delta_\phi = 0.62(2)$ is obtained, 
consistent with the CFT $\Delta_\phi=0.647$ in Tab.~\ref{tab:tab1}. {Such consistency between the finite size crossing point analysis of the SO(5) correlation ratio and the CFT spectrum, in giving rise to the same transition point and the scaling dimension of the SO(5) order parameter, supports the analysis employed in Ref.~\cite{chenPhases2023} and can be extended to the entire phase diagram around the multicritical point.}

{\noindent{\textcolor{blue}{\it Discussions.}}---}
{Following the pioneering work of Ref.~~\cite{zhouThe2023},}  our study provides a comprehensive
phase diagram for the (2+1)D SO(5) non-linear sigma
model with WZW term on a sphere, as well as 
the multicritical properties of the SO(5) transition. Our findings
reveal the {3D SO(5) multicriticality by identifying the CFT
tower structure when tuning irrelevant operators.} These results, combined with recent observations of the weakly first transition in the SU(2) spin models with N\'eel-VBS transtions~\cite{kuklovDeconfined2008,haradaPossibility2013,chenDeconfined2013,nahumDeconfined2015,zhaoMulticritical2020,songExtracting2023,song2023deconfined,takahashi2024so5}, provide a direction for the long-standing question on the existence of DQCP in various settings. 

One of the remaining questions is the nature of the symmetric phase and its previously determined non-Wilson-Fisher transition towards the VBS and N\'eel phases once one departs from the SO(5) line~\cite{chenPhases2023}. 
A few possible CFT scenarios have already been proposed~\cite{gazitConfinement2018,shackletonDeconfined2021,christosModel2023,christosDeconfined2024} that requires emergent $\mathbb{Z}_2$ or SU(2) gauge fields. We will present our analysis of these scenarios in a forthcoming work. {Also, the relation of the multicriticality and the previously discovered pseudocriticality at $V_1=0$~\cite{zhouThe2023} will be of interests to further explore}. In addition, conformal perturbation theory~\cite{Lao2023ConformalPerturbation} can also be applied to study the deviations from conformal spectrum in the future works.

\begin{acknowledgments}
{\noindent{\color{blue}\it{Acknowledgment.}}---} We thank Kai Sun, Yuxuan Wang, Meng Cheng, Cenke Xu, Wei Zhu, Yin-Chen He, Fakher Assaad, Anders Sandvik, Senthil Todadri, Slava Rychkov, Shai Chester, Subir Sachdev for valuable discussions on the related topic. We acknowledge the support from the Research Grants Council (RGC) of
Hong Kong Special Administrative Region of China (Project Nos. 17301721, AoE/P-701/20, 17309822, HKU C7037-
22GF, 17302223), the ANR/RGC Joint Research Scheme sponsored by RGC of Hong Kong and French National Research Agency (Project No. A HKU703/22), the GD-NSF (No. 2022A1515011007) and the HKU Seed Funding for
Strategic Interdisciplinary Research.
We thank HPC2021 system under the Information Technology Services and the Blackbody HPC system at the Department of Physics, University of Hong Kong, as well as the Beijng PARATERA Tech CO.,Ltd. (URL: https://cloud.paratera.com) for providing HPC resources that have contributed to the research results reported within this paper.
\end{acknowledgments}

\bibliography{bibtex}
\bibliographystyle{apsrev4-2}
%================================================================================
% Supplementary Information
% ================================================================================

\startsupplement

\begin{widetext}
\begin{center}
{\bf \uppercase{Supplemental Materials for \\[0.5em]
Emergent Conformal Symmetry at the Multicritical Point of (2+1)D SO(5) Model with Wess-Zumino-Witten Term on Sphere}}
\end{center}

\vskip3em

In Supplemental Materials \ref{sec:I}, we explain the spherical Landau level regularization of the SO(5) model. In \ref{sec:II}, we discuss the tensor representation of the SO(5) group, in which the irreducible representations with their quantum numbers and degeneracies are explicitly given. In \ref{sec:III}, we present further analysis of the energy spectra if other calibration criteria besides the $\Delta_{J^{\mu}}=2$ and $\Delta_{\mathcal{T}^{\mu\nu}}=3$ in the main text are used. The purpose of these alternative analysis is to explore the relation of thus obtained scaling dimensions with other recent related works such as Refs.~\cite{chesterBootstrapping2024,takahashi2024so5}.
In \ref{sec:IV}, we present the detailed derivation of the crossing point analysis of the SO(5) order parameter, employed in our previous work~\cite{chenPhases2023}.

\section{{Spherical Landau level regularization of SO(5) model}}
\label{sec:I}
\subsection{More on the SO(5) model}
Our notation is based on that used in Refs.~\cite{impoliteHalf2018,wangPhases2021,zhuUncovering2023,chenPhases2023}. 

We would like to project the SO(5) Hamiltonian onto the lowest Landau level (LLL) of the Haldane sphere. 
The original Hamiltonian is
\begin{equation}
H_\Gamma = \int d\mathbf{r}_1 d\mathbf{r}_2 U(\mathbf{r}_1,\mathbf{r}_2) 
[n^0(\mathbf{r}_1)n^0(\mathbf{r}_2)-u\sum_{i=1}^5 n^i(\mathbf{r}_1)n^i(\mathbf{r}_2)],
\end{equation}
where $n^i(\mathbf{r})={\bf\Psi(r)}^\dag\Gamma^i{\bf\Psi(r)}-\delta_{i0}$ is a local density operator with 
${\bf\Psi(r)}:=\psi_{\tau\sigma}(\mathbf{r})$ the 
4-component Dirac fermion annihilation operator with 
mixing valley $\tau$ and spin $\sigma$ indices.
And $\Gamma^0=\mathbb{I}\times\mathbb{I}$, $\Gamma^i=\{\tau_x\otimes\mathbb{I}, \tau_y\otimes\mathbb{I}, \tau_z\otimes\vec\sigma_x, \tau_z\otimes\vec\sigma_y, \tau_z\otimes\vec\sigma_z\}$ are the 5 mutually anticommuting matrices. 
Here, 
\begin{equation} \label{Eq:Interaction}
U(\mathbf{r}_1,\mathbf{r}_2) = \frac{g_0}{R^2}\delta(|\mathbf{r}_1-\mathbf{r}_2|)+
\frac{g_1}{R^4}\nabla^2\delta(|\mathbf{r}_1-\mathbf{r}_2|)
\end{equation}
describes the SO(5)-preserving short-range interactions with $g_0$ and $g_1$ terms.

\subsection{Spherical Landau level}
For electrons moving on the surface of a sphere with $4\pi s$ monopole ($2s\in Z$), the Hamiltonian is $H_0 = \tfrac{1}{2M_er^2}\Lambda_\mu^2$, and $\Lambda_\mu=\partial_\mu + \text{i}A_\mu$. 
The eigenstates are quantized into spherical Landau levels with 
energies $E_n = [n(n+1)+(2n+1)s]/(2M_er^2)$ and $n=0,1,\cdots$ the Landau level index. The $(n+1)_\mathrm{th}$ level is $(2s+2n+1)$-fold degenerate. We assume all interactions are much smaller than the energy gap between Landau levels, and just consider the lowest Landau level (LLL) $n=0$, which is $(2s+1)$-fold degenerate and we denote $N=2s+1$ as the system size of the problem. The wave-functions of LLL orbital are monopole harmonics 
\begin{equation}
\Phi_m(\theta,\phi) = N_m e^{im\phi}\cos^{s+m}(\tfrac{\theta}{2})\sin^{s-m}(\tfrac{\theta}{2}),
\end{equation}
with $m=-s,-s+1,\cdots,s$ and $N_m = \sqrt{\tfrac{(2s+1)!}{4\pi (s+m)!(s-m)!}}$.

\subsection{Details on the LLL projection}
The projection of $H_\Gamma$ on the LLL of the Haldane sphere is carried out as
\begin{eqnarray}
&& H_\Gamma^{(LLL)} =\int d\mathbf{r}_1d\mathbf{r}_2~U( {\mathbf{r}_1, \mathbf{r}_2})
\sum_{m_1,n_1}\Phi_{m_1}^\ast(\mathbf{r}_1)\Phi_{n_1}( \mathbf{r}_1)
\sum_{m_2,n_2}\Phi_{m_2}^\ast(\mathbf{r}_2)\Phi_{n_2}( \mathbf{r}_2) \times \\
&&( 
\sum_{\alpha}c_{m_1,\alpha}^\dag c_{n_1,\alpha} - 2\delta_{m_1,n_1} 
) 
( 
\sum_{\alpha}c_{m_2,\alpha}^\dag c_{n_2,\alpha} - 2\delta_{m_2,n_2} 
)
- u \sum_{i=1}^5 
( 
\sum_{\alpha,\beta}c_{m_1,\alpha}^\dag\Gamma_{\alpha,\beta}^i c_{n_1,\beta}
) 
( 
\sum_{\alpha,\beta}c_{m_2,\alpha}^\dag\Gamma_{\alpha,\beta}^i c_{n_2,\beta}
)
\end{eqnarray}

According to the Legendre polynomial $U\left( \left| {\mathbf{r}_{1} - \mathbf{r}_{2}} \right| \right) = {\sum\limits_{l = 0}^{\infty}{U_{l}P_{l}\left( {\cos\left( \Omega_{12} \right)} \right)}} = {\sum\limits_{l}{U_{l}\frac{4\pi}{2l+ 1}{\sum\limits_{m = - l}^{l}{Y_{l,m}^{*}\left( \Omega_{1} \right)Y_{l,m}\left( \Omega_{2} \right)}}}}$. 
We then arrive at the form, 
\begin{align}
H_\Gamma^{(LLL)} =&\sum_{l,m} U_l\sum_{m_1,m_2} (-1)^{2s+m+m_1+m_2}\tfrac{(2s+1)^2}{2}  
\begin{pmatrix}s&l&s\\-m_1&-m&m_1+m\end{pmatrix}
\begin{pmatrix}s&l&s\\-m_2&m&m_2-m\end{pmatrix}
\begin{pmatrix}s&l&s\\-s&0&s\end{pmatrix}^2 \nonumber\\
&\times( 
\sum_{\alpha}c_{m_1,\alpha}^\dag c_{n_1,\alpha} - 2\delta_{m_1,n_1} 
) 
( 
\sum_{\alpha}c_{m_2,\alpha}^\dag c_{n_2,\alpha} - 2\delta_{m_2,n_2} 
)
- u \sum_{i=1}^5 
( 
\sum_{\alpha,\beta}c_{m_1,\alpha}^\dag\Gamma_{\alpha,\beta}^i c_{n_1,\beta}
) 
( 
\sum_{\alpha,\beta}c_{m_2,\alpha}^\dag\Gamma_{\alpha,\beta}^i c_{n_2,\beta}
)\nonumber\\
=& \sum_{m_1,m_2,m} V_{m_1,m_2,m_2-m,m_1+m} \nonumber\\
&\times( 
\sum_{\alpha}c_{m_1,\alpha}^\dag c_{n_1,\alpha} - 2\delta_{m_1,n_1} 
) 
( 
\sum_{\alpha}c_{m_2,\alpha}^\dag c_{n_2,\alpha} - 2\delta_{m_2,n_2} 
)
- u \sum_{i=1}^5 
( 
\sum_{\alpha,\beta}c_{m_1,\alpha}^\dag\Gamma_{\alpha,\beta}^i c_{n_1,\beta}
) 
( 
\sum_{\alpha,\beta}c_{m_2,\alpha}^\dag\Gamma_{\alpha,\beta}^i c_{n_2,\beta}
)
\end{align}
with 
\begin{equation} \label{Eq:FormFactor}
V_{m_1,m_2,m_3,m_4} = (-1)^{2s+m_1+2m_2-m_3}\tfrac{(2s+1)^2}{2} \sum_l U_l (2l+1)  
\begin{pmatrix}s&l&s\\-m_1&m_1-m_4&m_4\end{pmatrix}
\begin{pmatrix}s&l&s\\-m_2&m_2-m_3&m_3\end{pmatrix}
\begin{pmatrix}s&l&s\\-s&0&s\end{pmatrix}^2. 
\end{equation}

\subsection{Connection to Haldane's pseudo-potential}
In this part, we discussion the interaction we considered [c.f. \Eq{Eq:Interaction}] and 
its relation with the Haldane's pseudopotential $V_l$. The form factor
$V_{m_1,m_2,m_3,m_4}$ [c.f. \Eq{Eq:FormFactor} ] is connected to $V_l$ by 
\begin{equation}
V_{m_1,m_2,m_3,m_4} = \sum_l V_l (4s-2l+1) 
\begin{pmatrix}s&s&2s-l\\m_1&m_2&-m_1-m_2\end{pmatrix}
\begin{pmatrix}s&s&2s-l\\m_4&m_3&-m_4-m_3\end{pmatrix}.
\end{equation}
The relation between $U_l$ and $V_l$ is then given by, 
\begin{equation}
V_{2s-l} = (-)^{2s+l}\sum_k U_k 
\begin{pmatrix}s&k&s\\-s&0&s\end{pmatrix}^2
\begin{Bmatrix}s&s&l\\s&s&k\end{Bmatrix}.
\end{equation}

To be specific, for the short-range interaction 
$
U(\mathbf{r}_1,\mathbf{r}_2) = \frac{g_0}{R^2}\delta(|\mathbf{r}_1-\mathbf{r}_2|)+
\frac{g_1}{R^4}\nabla^2\delta(|\mathbf{r}_1-\mathbf{r}_2|)$, only the contact interaction 
of the form of delta function and its derivatives are considered, 
and only $V_0$ and $V_1$ are involved in the Haldane's pseudopotential. 
And the relations between $(g_0, g_1)$ and $(V_0, V_1)$ is given by
\begin{equation}
V_l = 
\begin{cases}
\frac{g_0(2s+1)-g_1s}{4s+1}, &\text{if $l=0$}\\
\frac{g_1s}{4s-1}, &\text{if $l=1$}.\\
\end{cases}
\end{equation}

\section{Tensor representation of SO($N$) group}
\label{sec:II}

In this section, we briefly recapitulate basic conclusions of SO($N$) group 
and its irreducible representation~\cite{ZeeGroup,Sternberg1994Group}. 
These information forms the foundation upon which we classify the energy levels in the ED and DMRG simulations of the SO(5) model with WZW term.

By definition, the SO($N$) group is the group of all 
$N\times N$ matrices $R$ satisfying $R^T R = I $ (orthogonal condition) and $\det(R) = 1$. 
These $N\times N$ matrices thus form the defining representation, furnished by $N$-dimensional 
vectors.

To systematically construct representations with larger dimensions, one needs more complicated objects 
than vectors, i.e., a rank-$j$ tensor denotes as $T^{\mu_1\mu_2\cdots\mu_j}$ with $j$ indices, 
where $\mu_j$ can take the $N$ integers. It transforms as 
$$ \tilde T^{\nu_1\nu_2\cdots\nu_j} = R^{\nu_1}_{\mu_1}R^{\nu_2}_{\mu_2}\cdots R^{\nu_j}_{\mu_j} ~T^{\mu_1\mu_2\cdots\mu_j}.$$
The number of elements in rank-$j$ tensor $T^{\mu_1\mu_2\cdots\mu_j}$ is $N^j$, and such generic 
rank-$j$ tensors thus furnish a $N^j\times N^j$ representation, which generally speaking is reducible. To be specific, it can be decomposed into sectors with different index symmetries.

For instance, a rank-2 tensor can be decomposed as the symmetric part and 
the anti-symmetric part, i.e., $T^{ij} = S^{ij} + A^{ij}$, 
where $S^{ij}=(T^{ij}+T^{ji})/2$ is a symmetric tensor and 
$A^{ij} = (T^{ij}-T^{ij})/2$ is a anti-symmetric tensor. 
It is straightforward to check such index permutation symmetries are preserved 
under SO($N$) transformations. That is, a subset of tensors with a given index symmetry 
constitute an invariant subspace. Note that, the trace of a tensor remains the same under 
SO($N$) rotation, which means that, the symmetric tensor can be further decomposed into 
the traceless symmetric tensor and the trace itself. To sum up, for rank-2 tensors, 
the $N^2$-dimensional space can be decomposed into, a $(N(N+1)/2-1)$-dimensional invariant 
subspace (furnished by the traceless symmetric tensors), a $N(N-1)/2$-dimensional invariant 
subspace (furnished by the anti-symmetric tensors), and a one-dimensional invariant subspace 
(by the trace of the rank-2 tensors), denoted as, 
$$ N^2 = \tfrac{N(N+1)}{2}-1 ~\oplus~ \tfrac{N(N-1)}{2} ~\oplus~ 1.$$ 
For SO(5) group, this means 
$$25 = 14 \oplus 10 \oplus 1.$$

For more complicated symmetries in higher rank tensors, one usually resorts to the simple yet 
powerful tools of the Young tableaux, where each tableau represents a specific process of 
symmetrization and anti-symmetrization for the indices of the tensor 
$T^{\mu_1\mu_2\cdots\mu_j}$~\cite{Sternberg1994Group}. In practice, one first fill in all indices into a given Young 
diagram [$\lambda$], then symmetrizes all indices for each row and anti-symmetrizes all indices for each 
column. For instance, the Young diagram $[2,1]$, depicted as 
\begin{Young}
$\mu$ & $\rho$ \cr
$\nu$ \cr
\end{Young},
represents the mix-symmetry rank-3 tensors $M^{\mu\rho\nu} = (T^{\nu\mu\rho}+T^{\nu\rho\mu}) - (T^{\mu\nu\rho}+T^{\mu\rho\nu})$. 
The dimension of a tensor representation can be given by the Young tableaux theory. 
When the number of rows in a given Young diagram [$\lambda$] is not equal to $N/2$ (applied to the considered 
$N=5$ case here), the dimension is then given by the ratio of two tableaux $\frac{Y^{[\lambda]}_T}{Y^{[
\lambda]}_h}$. 
Here the denominator tableaux $Y^{[\lambda]}_h$ is simply given by filling in the hook length $h_{ij}$ into the 
box of the $i$-th row and $j$-th column, and calculating their product. 
The numerator tableaux $Y^{[\lambda]}_T$ is slightly more complicated, where for a given box the number 
filled is the sum of that in the corresponding box of a set of tableaux $\{Y^{[\lambda]}_{T_a}\}$. 
The rules to write down the series of $Y^{[\lambda]}_{T_a}$ are as follows.
\begin{enumerate}
\item Tableaux $Y^{[\lambda]}_{T_0}$ is given by filling $N-i+j$ into the box at the $i$-th row 
and $j$-th column;
\item Let $[\lambda^1] = [\lambda]$. We then define a series of diagram $[\lambda^a]$ by removing 
the first row and column of the diagram $[\lambda^{a-1}]$, until the number of columns is less than $2$.
\item Given diagram $[\lambda^a]$, for the first-$r$ boxes (with $r$ being the row of $[\lambda^a]$) 
of the hook $(1,1)$, we successively fill in 
$(\lambda^a_1-1), (\lambda^a_2-1),\cdots,(\lambda^a_r-1)$. For each hook ${(i,1)}$ 
with $1\leq i\leq r$, we fill $(-1)$ into the last-$(\lambda^a_i -1)$ boxes. 
\end{enumerate}
This ends the definition of tableaux $Y^{[\lambda]}_{T_a}$. 

For the SO(5) group we considered in the main text, a few of the lowest-dimensional 
representations (without spinor) are listed as follows.

\begin{enumerate}
\item $\begin{Young} \cr \end{Young}$ (rank-1 tensor, the vector representation): the dimension is 
$\frac{\ytableausetup{smalltableaux}\ytableaushort{5}}{\ytableausetup{smalltableaux}\ytableaushort{1}}
=\frac{5}{1}=5$;

\item $\begin{Young} & \cr \end{Young}$ (the traceless symmetric rank-2 tensor): the dimension is 
$\frac{\ytableausetup{smalltableaux,centertableaux}\ytableaushort{56}+\ytableaushort{{\text{-}1}1}}{\ytableausetup{smalltableaux}\ytableaushort{21}} = 
\frac{\ytableausetup{smalltableaux}\ytableaushort{47}}{\ytableausetup{smalltableaux}\ytableaushort{21}} 
=\frac{4\times7}{2\times1}=14$;

\item $\begin{Young} \cr \cr \end{Young}$ (the anti-symmetric rank-2 tensor): the dimension is 
$\frac{\ytableausetup{smalltableaux}\ytableaushort{5,4}}{\ytableausetup{smalltableaux}\ytableaushort{2,1}}
=\frac{5\times4}{2\times1}=10$;

\item $\begin{Young} &&\cr \end{Young}$ (the fully symmetric rank-3 tensor): the dimension is 
$\frac{\ytableausetup{smalltableaux,centertableaux}\ytableaushort{567}+\ytableaushort{{\text{-}1}{\text{-}1}2}}{\ytableausetup{smalltableaux}\ytableaushort{321}} = 
\frac{\ytableausetup{smalltableaux}\ytableaushort{459}}{\ytableausetup{smalltableaux}\ytableaushort{321}} 
=\frac{4\times5\times9}{3\times2\times1}=30$; 

\item $\begin{Young} &\cr \cr \end{Young}$ (the mixed-symmetry rank-3 tensor): the dimension is 
$\frac{\ytableausetup{smalltableaux,centertableaux}\ytableaushort{56,4}+\ytableaushort{{0}1,{\text{-}1}}}{\ytableausetup{smalltableaux}\ytableaushort{31,1}} = 
\frac{\ytableausetup{smalltableaux}\ytableaushort{57,3}}{\ytableausetup{smalltableaux}\ytableaushort{31,1}} 
=\frac{5\times7\times3}{3\times1\times1}=35$; 

\item etc.
\end{enumerate}

In our ED calculation, we have the conserved quantum numbers, 
the $z$-component of the total spin $\sigma^z$, the total valley $\tau^z$, the total charge number 
$q^z$. The tensor representations manifest themselves through the degeneracies in each 
$(\sigma^z, \tau^z, q^z)$ sector. As shown in \Tab{tab:tabs1} we list the SO(5) representations and their corresponding 
degeneracies in $(\sigma^z, \tau^z, q^z=2(2s+1))$ sectors. 
Here, we only list the sectors with $\sigma^z\geq0$ and $\tau^z\geq0$, and 
the sectors $(\pm\sigma^z, \pm\tau^z)$ should have the same degeneracy.

\begin{table}[h!]
\newcolumntype{M}[1]{>{\centering\arraybackslash}m{#1}}
\ytableausetup{boxsize=.55em}
\caption{The Young diagrams of different SO(5) irreducible representations (denoted as IREP) 
and the corresponding state degeneracies in different $(\sigma^z,\tau^z)$ sectors. 
}
\begin{tabular}{M{1.5cm}|M{1.5cm}|M{1cm}M{1cm}M{1cm}M{1cm}M{1cm}M{1cm}M{1cm}M{1cm}M{1cm}M{1cm}}
    \toprule
    SO(5) IREP&Young diagram&(0,0)&(0,2)&(0,4)&(0,6)&(2,0)&(2,2)&(2,4)&(4,0)&(4,2)&(6,0) 
    \\[4pt]
    \hline
     \bf1  &              &1& & & & & & & & &  \\[6pt]
     \bf5  &\ydiagram{1}  &1&1& & &1& & & & &  \\[6pt]
     \bf10 &\ydiagram{1,1}&2&1& & &1&1& & & &  \\[6pt]
     \bf14 &\ydiagram{2}  &2&1&1& &1&1& &1& &  \\[6pt]
     \bf30 &\ydiagram{3}  &2&2&1&1&2&1&1&1&1&1 \\[6pt]
     \bf35 &\ydiagram{2,1}&3&3&2& &3&2&1&1&1&  \\[6pt]
     \hline\hline
\end{tabular}
\label{tab:tabs1}
\end{table}

In our DMRG calculation, we have the conserved quantum numbers, 
the total spin $S$, the total charge number 
$q^z$. The tensor representations manifest themselves through the degeneracies in each 
$(S, q^z)$ sector. As shown in \Tab{tab:tabs2} we list the SO(5) representations and their corresponding 
degeneracies in $(S, q^z=2(2s+1))$ sectors. 

\begin{table}[h!]
\newcolumntype{M}[1]{>{\centering\arraybackslash}m{#1}}
\ytableausetup{boxsize=.55em}
\caption{The Young diagrams of different SO(5) irreducible representations (denoted as IREP) 
and the corresponding state degeneracies in different sectors with total spin $S$ 
at half-filling case $q^z = 2(2s+1)$. 
}
\begin{tabular}{M{1.5cm}|M{1.5cm}|M{1cm}M{1cm}M{1cm}M{1cm}}
    \toprule
    SO(5) IREP&Young diagram&0&1&2&3 
    \\[4pt]
    \hline
     \bf1  &              &1& & &   \\[6pt]
     \bf5  &\ydiagram{1}  &2&3& &   \\[6pt]
     \bf10 &\ydiagram{1,1}&1&9& &   \\[6pt]
     \bf14 &\ydiagram{2}  &3&6&5&  \\[6pt]
     \bf30 &\ydiagram{3}  &4&9&10 \\[6pt]
     \bf35 &\ydiagram{2,1}&2&18&15  \\[6pt]
     \hline\hline
\end{tabular}
\label{tab:tabs2}
\end{table}

\section{Analysis of Recalibrated Spectrum}
\label{sec:III}
In the main text, we use the criteria of $\Delta_{J^{\mu}}=2$ and $\Delta_{\mathcal{T}^{\mu\nu}}=3$ to determine the microscopic parameter at which the CFT energy level structure is revealed. As shown in \Tab{tab:tabs3}, we list our obtained scaling dimensions of 
several operators, namely, the order parameter $\Delta_\phi$, singlet operator $\Delta_S$, 
the lowest rank-2 tensor $\Delta_T$, and the symmetry current $\Delta_{J^\mu}$, together with 
the recent conformal bootstrap data assuming tricritical point~\cite{chesterBootstrapping2024}, 
a recent ED study of the same SO(5) model we consider in fuzzy sphere~\cite{zhouThe2023},
and a recent QMC study of $J$-$Q$ model~\cite{takahashi2024so5}.

\begin{table}[h!]
\newcolumntype{M}[1]{>{\centering\arraybackslash}m{#1}}
\caption{Scaling dimension of relevant primary operator at $V_1=0.3$ and those from related works.}
\begin{tabular}{c|M{1.5cm}M{1.5cm}M{1.5cm}M{1.5cm}}
    \toprule
    Operators & $\phi$ & $S$ & $T$ & $J^\mu$  \\
    \hline
    This work & 0.646$^{\hspace{1.3mm}}$  & 2.884 & 1.633  & 2.000$^\ast$  \\       
    Conformal Bootstrap tricritical point~\cite{chesterBootstrapping2024} & 0.630${^\ast}$  & 2.359 & 1.519  & 2.000${^\ast}$   \\    
    Fuzzy sphere with SO(5) symmetry only and $V_1=0$~\cite{zhouThe2023}  & 0.585$^{\hspace{1.3mm}}$  & 2.831 & 1.458 & 2.000$^\ast$   \\    
        $J$-$Q$ model~\cite{takahashi2024so5} & 0.607(4) & 2.273(4) & 1.417(7) & 2.01(3) \\      
    \hline\hline
\end{tabular}
\label{tab:tabs3}
\end{table}

As shown in the main text, the analysis still has noticeable finite size effect even 
if we tune the length of the interaction. In this section, we try to loose the 
constraints of 
$\Delta_{J^{\mu}}=2$ and $\Delta_{\mathcal{T}^{\mu\nu}}=3$ on the finite size data, 
and test our energy levels with other criteria in such a way that, 
given the recent relevant work from 
the conformal bootstrap analysis of the deconfined quantum tricrticality~\cite{chesterBootstrapping2024}, 
and Quantum Monte Carlo simulation of $J$-$Q$ model~\cite{takahashi2024so5}, 
how large is the degree of freedom our obtained scaling dimensions 
are consistent with these other results.

As shown in \Tab{tab:tabs4}, for the cases of $V_1=0.3$, 
we instead calibrate our energy spectrum to make the scaling dimension of symmetry current 
$\Delta_J = 2$ and order parameter $\Delta_\phi =0.630$ (the one taken in the conformal bootstrap work
\cite{chesterBootstrapping2024}).
It can be seen that, when $V_1=0.3$, the rank-2 tensor has a scaling dimension of $1.593$ which is closer to the value of $1.519$ given by the conformal bootstrap, 
whereas the singlet operator is $2.774$ which is far from $2.359$. 

\begin{table}[h!]
\caption{Scaling dimensions of relevant primary operators at $V_1=0.3$ for various system sizes $N$ when recalibrating the operator spectrum to make $\Delta_J=2$ and $\Delta_\phi=0.630$ (the one taken in conformal bootstrap~\cite{chesterBootstrapping2024}).}
\begin{tabular}{c|ccccc}
    \toprule
         & &  & Op. &  &  \\
    \hline
     & \hsbc{$\phi^{~~}$} & \hsbc{$T^{~~}$} & \hsbc{$S^{~~}$} & \hsbc{$J^\mu~$} & \hsbc{$\mathcal{T}^{\mu\nu}$} \\
    \hline
    $N=4$ & 0.630$^{\ast}$  & 1.590 & 2.777  & 2.000$^\ast$  & 3.047  \\       
    $N=5$ & 0.630$^{\ast}$  & 1.592 & 2.774  & 2.000$^\ast$  & 3.042  \\       
    $N=6$ & 0.630$^{\ast}$  & 1.593 & 2.780  & 2.000$^\ast$  & 3.046  \\       
    Conformal Bootstrap tricritical point~\cite{chesterBootstrapping2024} & 0.630$^\ast$ & 1.519 & 2.359 & 2.000$^\ast$ & --- \\      
    \hline\hline
\end{tabular}
\label{tab:tabs4}
\end{table}

\begin{table}[h!]
\caption{Scaling dimensions of relevant primary operators at $V_1=0.3$ for various system sizes $N$ when recalibrating the operator spectrum to make $\Delta_J=2$ and $\Delta_\phi=0.607$ (the one obtained in $J$-$Q$ model~\cite{takahashi2024so5}).}
\begin{tabular}{c|ccccc}
    \toprule
         & &  & Op. &  &  \\
    \hline
     & \hsbc{$\phi^{~~}$} & \hsbc{$T^{~~}$} & \hsbc{$S^{~~}$} & \hsbc{$J^\mu~$} & \hsbc{$\mathcal{T}^{\mu\nu}$} \\
    \hline
    $N=4$ & 0.607$^{\ast}$  & 1.531 & 2.688  & 2.000$^\ast$  & 3.133  \\       
    $N=5$ & 0.607$^{\ast}$  & 1.533 & 2.681  & 2.000$^\ast$  & 3.121  \\       
    $N=6$ & 0.607$^{\ast}$  & 1.533 & 2.684  & 2.000$^\ast$  & 3.121  \\       
    $J$-$Q$ model~\cite{takahashi2024so5} & 0.607(4) & 1.417(7) & 2.273(4) & 2.01(3) & --- \\      
    \hline\hline
\end{tabular}
\label{tab:tabs5}
\end{table}

As shown in \Tab{tab:tabs5}, for the cases of $V_1=0.3$, 
we also calibrate our energy spectrum to make the scaling dimension of symmetry current 
$\Delta_J = 2$ and order parameter $\Delta_\phi =0.607$ (the one obtained in the $J$-$Q$ model
\cite{takahashi2024so5}).
It can be seen that, when $V_1=0.3$, the rank-2 tensor has a scaling dimension of $1.533$ which is closer to the value of $1.417$ obtained in the $J$-$Q$ model, 
whereas the singlet operator is $2.684$ which is far from $2.273$. 

Based on these analyses, we conclude that all these recent works on existence of the multicritical point  of the SO(5), although still differ at the second/third significant digit, but they reveal the consistent picture that for the SO(5) CFT we find in this work, there exists two relevant singlets, the $\Delta_{S}$ and $\Delta_{T}$, whose scaling dimensions are both smaller than 3. And by tuning $V_1=0.3$ and calibrate with 
$\Delta_{\phi}$ taken in the conformal bootstrap work~\cite{chesterBootstrapping2024} and obtained
in the QMC study of $J$-$Q$ model~\cite{takahashi2024so5}, 
we observe the $\Delta_{s}$ become closer to the bootstrap value and the $J$-$Q$ value. 
Such analysis supports the conclusion that the emergent CFT we have discovered, 
along with those from other related works, is a multicritical point.

\section{Crossing point analysis}
\label{sec:IV}

In this section, we provide the detailed derivation for the scaling form of 
the crossing points, such that the position of the critical point and the associated 
scaling dimension can be obtained in a controlled manner from the finite size data. 
Such crossing point analysis has been widely applied and tested for quantum criticality of 
2D Ising, SU(2) and other spin models~\cite{qinDuality2017,shaoQuantum2016,maAnomalous2018,maRole2019} 
and can be further traced back to Fisher's ``phenomenological renormalization'', 
which was first numerically tested with transfer matrix results for 
the Ising model in Ref.~\cite{luckCorrections1985}. It has been used in our previous work~\cite{chenPhases2023} to determine the SO(5) phase diagram.

Let's consider the standard form of finite-size scaling 
for an arbitrary observable, 
\begin{equation}
O(\delta, L) = L^{-\kappa/\nu} f(\delta L^{1/\nu}, \lambda L^{-\omega}).
\end{equation}
Here, $\delta=q-q_c$ is the deviation from the transition point $q_c$, and we also 
consider the correction from the leading irrelevant field $\lambda$ and its 
corresponding exponent $\omega$. 
In practise, due to the limit of computational resources, we only increase the system size
by $x$, and consider the crossing point of observable between size pair $(N, N+x)$. 
For the sake of notation simplicity, we express the scaling form as a function of total number of 
size $N$ instead of linear size $L\sim\sqrt{N}$, i.e., 
\begin{equation}
O(\delta, N) = N^{-\frac{\kappa}{2\nu}} f(\delta N^{\frac{1}{2\nu}}, \lambda N^{-\frac{\omega}{2}}) 
= N^{-\frac{\kappa}{2\nu}} (a_0 + a_1 \delta N^{\frac{1}{2\nu}} + b_1 N^{-\frac{\omega}{2}} + \cdots), 
\end{equation}
where the second equality relation is simply from Taylor's expansion up to first-order. 
Similarly, for system size $N+x$, we have 
\begin{equation}
O(\delta, N+x) 
= (N+x)^{-\frac{\kappa}{2\nu}} (a_0 + a_1 \delta (N+x)^{\frac{1}{2\nu}} 
+ b_1 (N+x)^{-\frac{\omega}{2}} + \cdots). 
\end{equation}
Then, at the crossing point $\delta^\ast$, by definition we have 
$O(\delta^\ast,N) = O(\delta^\ast,N+x)$, which leads to the scaling form 
for the crossing point itself and the observable at the crossing point, 
\begin{equation}
\delta^\ast(N) = \frac{a_0}{a_1} \frac{(1+x/N)^{-\frac{\kappa}{2\nu}}-1}{1-(1+x/N)^{\frac{1-\kappa}{2\nu}}} N^{-\frac{1}{2\nu}} + \frac{b_1}{a_1} \frac{(1+x/N)^{-\frac{\omega}{2}-\frac{\kappa}{2\nu}}-1}{1-(1+x/N)^{\frac{1-\kappa}{2\nu}}} N^{-\frac{1}{2\nu}-\frac{\omega}{2}} + \cdots.
\end{equation}

In the case of Binder ratio, we have $\kappa=0$ and when $x\ll N$, we then arrive at 
\begin{equation}\label{Eq:CrossingScaling}
\delta^\ast(N) = a N^{-\frac{1}{2\nu}-\frac{\omega}{2}} + \cdots.
\end{equation}

From the above scaling form~[c.f. \Eq{Eq:CrossingScaling}], in principle, 
we can fit the finite-size data of the crossing point, 
from which the critical point can be extracted. 
In the main text, we applied this method to determine the transition point between 
the SO(5) symmetry-breaking phase and the symmetric phase.

To determine the scaling dimension of order parameter $\Delta_\phi$, 
by considering the order parameter 
$\phi(\delta, N) = N^{-\frac{\Delta_\phi}{2}}(a_0 + a_1\delta N^{\frac{1}{2\nu}}+\cdots)$, 
the logarithmic of the order parameter between size pair $(N,N+x)$ will be 
\begin{equation}
\log{\frac{\phi(\delta,N+x)}{\phi(\delta,N)}} =  -\frac{\Delta_\phi}{2}\frac{x}{N} + \frac{a_1}{a_0}\delta N^{\frac{1}{2\nu}}\frac{x}{N}.
\end{equation}

Then, we can define the finite-size value of $\Delta^\ast(N)$ as
\begin{equation}
\Delta_\phi^\ast(N)\equiv -\frac{2N}{x} \log\frac{\phi(\delta,N+x)}{\phi(\delta,N)},
\end{equation} 
which will take the scaling form as 
\begin{equation}
\Delta_\phi^\ast(N) = \Delta_\phi + c N^{\frac{1}{2\nu}},
\end{equation}
where both of the sides will take the value of scaling dimension $\Delta_\phi$ when 
$N$ is extrapolated to $0$.

\end{widetext}
\end{document}